\documentclass[floatfix]{emulateapj}
\usepackage{amssymb}
\usepackage{amsmath}
\usepackage{graphicx}
\usepackage{natbib}
\usepackage{color}
\usepackage{txfonts}
\newcommand{\unit}[1]{\,\text{#1}}
\newcommand{\Msun}{\hbox{$\,$M$_\odot$} }

\newcommand{\Teff}{\hbox{$\,T_{\rm eff}$} }
\newcommand{\K}{{\rm K}}
\newcommand{\logg}{{\log(g)}}
\newcommand{\jcap}{J. Cosmology Astropart. Phys.}

\shorttitle{Constraints on Dark Stars from the EBL}
\shortauthors{Maurer et al.}

\begin{document}
\title{Dark matter powered stars: Constraints from the extragalactic background light}

\author{A. Maurer, M. Raue, T. Kneiske, and D. Horns}
\affil{Institut f\"ur Experimentalphysik, Universit\"at Hamburg,
              Luruper Chaussee 149, D-22761 Hamburg}
\email{andreas.maurer@physik.uni-hamburg.de}

\author{D. Els\"asser}
\affil{Institut f\"ur Theoretische Physik und Astrophysik, Am Hubland, D-97074 W\"urzburg}

\and

\author{P. H. Hauschildt}
\affil{Hamburger Sternwarte, Gojenbergsweg 112, D-21029 Hamburg}

\begin{abstract}
The existence of predominantly cold non-baryonic dark matter is unambiguously demonstrated by several observations (e.g., structure formation, big bang nucleosynthesis, gravitational lensing, and rotational curves of spiral galaxies).
A candidate well motivated by particle physics is a weakly interacting massive particle (WIMP). Self-annihilating WIMPs would affect the stellar evolution especially in the early universe. Stars powered by self-annihilating WIMP dark matter should possess different properties compared with standard stars. While a direct detection of such dark matter powered stars seems very challenging, their cumulative emission might leave an imprint in the diffuse metagalactic radiation fields, in particular in the mid-infrared part of the electromagnetic spectrum. In this work the possible contributions of dark matter powered stars (dark stars; DSs) to the extragalactic background light (EBL) are calculated. It is shown that existing data and limits of the EBL intensity can already be used to rule out some DS parameter sets.
\end{abstract}



\section{Introduction}
    Several independent observations provide compelling evidence for an accelerated expanding universe with a matter content dominated by a non-luminous (``dark'') component ($\Lambda$CDM cosmology). Relevant parameters for this so-called $\Lambda$CDM cosmology have been measured with high accuracy with WMAP \citep{komatsu:2011a} and baryonic acoustic oscillations surveys and will be further refined with upcoming Planck data.
    Large sky surveys (e.g., SDSS \citealt{abazajian:2009a}, 2dF \citealt{colless:2003a}) and numerical simulations like, e.g., the Millennium Run \citep{springel:2005a} point toward a convincing picture of large-scale structure formation within the cosmological concordance model.
    A promising particle candidate for the dark matter (DM) content of the universe is the WIMP (weakly interacting massive particle; see e.g., \citealt{jungman:1996a, bertone:2005a} for a review article). Such particles can naturally have annihilation or decay channels leading to standard model particles.

    The epoch where the first stars form ($z \sim 30$) is not yet observable
    with today's astronomical instruments. The circumstances and mechanisms of stellar genesis are 
    still topics of ongoing analysis and rely on sophisticated numerical simulations (for a review see
    ,e.g., \citealt{bromm:2004a}).
    Recently, studies discussed the impact of WIMP dark matter on the formation of the first stars 
    \citep{spolyar:2008a, iocco:2008a, iocco:2008b, freese:2008b}. During the epoch of reionization of the universe, DM may have affected the formation and evolution of stars. Assuming that
    self-annihilating particles provide the dark matter content of the universe, this new source of energy injection into the first stars may alter their properties.
    The energy injection from self-annihilating WIMPs can delay or even prevent the nuclear
    hydrogen burning.

 In principle,
two mechanisms could lead to dark matter accretion into a star. The first one, investigated by \citet{spolyar:2008a}, is adiabatic contraction (AC) where 
additional dark matter from outside the forming first star is gravitationally pulled along with
accreting baryonic gas onto the proto-star. Due to the very low surface temperature of
the Dark Star compared to its enormous mass there should be no or very little
radiative feedback mechanisms preventing further accretion. The other possibility
to replenish the dark matter inside a star is capture of WIMPs via scattering. The efficiency of this mechanism is depends on the product of the elastic scattering cross section between baryonic and dark matter
and the surrounding dark matter density \citep{iocco:2008a, iocco:2008b,freese:2008b}.
Both mechanisms can lead to generic properties of the DS: low surface temperatures ($\sim 5\,000 - 10\,000 \text{ K}$), high luminosities ($\sim 10^6 \text{ L}_{\odot}$) and presumably longer lifetimes than conventional Pop III stars.   
After a ``dark phase'', the star evolves as a normal zero-age main-sequence star. 

Direct detection, even of a Dark Star cluster, is a challenging task which may only be possible with future instruments like the upcoming James Webb Space Telescope (JWST; \citealt{gardner:2006a}) and under the assumption of an optimistic model \citep{zackrisson:2010a}. A different approach for the search of emission in the early universe is to probe the diffuse metagalactic radiation field (MRF), see, e.g., \citet{raue:2009a}. The optical to infrared part of the local MRF is also known as extragalactic background light (EBL; for a review see, e.g., \citealt{hauser:2001a}).
 Its main contribution comes from integrated starlight and thermal dust emissions of all cosmic epochs \citep{kneiske:2002a, kneiske:2004a, stecker:2006a, franceschini:2008a, primack:2008a, gilmore:2009a, finke:2010a, kneiske:2010a}. This fact makes the EBL a unique probe for the integrated star formation history of the universe.
There are different types of observational approaches to measure the EBL. Direct observations e.g., with the DIRBE instrument on board the COBE satellite suffer from prominent foreground emission, like zodiacal light, that is caused by scattered sunlight by dust in the zodiacal cloud, and diffuse galactic radiation \citep{hauser:1998a}. 
Lower limits to the EBL are derived from integrated galaxy number counts which are available up to a redshift $\approx 2$ from Hubble Space Telescope \citep{madau:2000a} and the Spitzer instrument \citep{fazio:2004a}. 
A powerful method for obtaining upper limits on the EBL density makes use of the spectra from very high energy (VHE) $\gamma$-ray sources, especially blazars (see, e.g., \citealt{Mazin:2007a}).  
Galaxy number counts and upper limits from VHE $\gamma$-ray observations can be used to constrain possible DS scenarios.

 In this paper, the so far unknown contribution of DS to the EBL is calculated. Constraints for some DS scenarios are derived as well as a convenient parameterization to calculate the maximum EBL density produced by a set of DS parameter values.
The paper is organized as follows:
In Sect. \ref{sec:method} the model calculations for the EBL from DSs are described. In Sect. \ref{sec:results} the resulting EBL for different sets of DS parameters is calculated and compared with recent data and EBL models. The multidimensional DS parameter space is constrained and a parameterization for the peak EBL contribution of DS is derived. Sect. \ref{sec:discussion} summarizes the obtained results and compares them with existing direct and indirect approaches to detect/constrain DS.
   
Throughout this paper a flat Friedmann cosmology is adopted with $\Omega_{m} = 0.3$, $\Omega_{\Lambda} = 0.7$ and a Hubble constant of $H_0 = 70 \unit{km} \unit{s}^{-1} \unit{Mpc}^{-1}$.

\section{The DS contribution to the extragalactic background light (EBL)}
\label{sec:method}
In order to calculate the EBL produced by dark matter burning stars a forward evolution model is used based
on the calculations of \citet{kneiske:2002a} and \citet{raue:2009a} (see also \citealt{dwek:1998a,salamon:1998a}). 
In the model, the specific intensity of the EBL $I_{\nu}(z)$ is obtained by integrating 
the specific comoving luminosity density $\varepsilon_{\nu}(z)$ over redshift $z$
\begin{equation}
  I_{\nu}(z)=\frac{c}{4\pi}\int\limits_z^{z_{\text{max}}}\text{d}z'\,\varepsilon_{\nu'}(z')\left|\frac{\text{d}t}{\text{d}z'}\right| 
  \text{\label{formula:EBL}}
\end{equation}
where $\varepsilon_{\nu}(z)$ is given by
\begin{equation}
\varepsilon_{\nu}(z) = \int\limits_z^{z_{\text{max}}}\text{d}z'\,L_{\nu}(t(z)-t(z')) \dot{\rho}_{\ast}(z')\left|\frac{\text{d}t}{\text{d}z'}\right| \, .
\text{\label{formula:emissivity}}
\end{equation}
$L_{\nu}(t)$ is the time-dependent specific luminosity, $\dot{\rho}_{\ast}(z)$ the comoving formation rate of
DSs at a given redshift $z$, and $\nu'=\nu (1+z)/(1+z')$ is the redshifted frequency. The integration limit $z_{\text{max}}$ determines the maximum redshift
where dark matter burning stars begin to form.
Cosmological parameters enter through
\begin{align}
 \left|\frac{\text{d}t}{\text{d}z}\right| &= \frac{1}{H_0(1+z)E(z)} \\
 E(z)^2 &= \Omega_r(1+z)^4+\Omega_m(1+z)^3+\Omega_k(1+z)^2+\Omega_{\Lambda} ,
\end{align}
as described in, e.g., \citet{peebles:1993a}.
Further details on the method and formulae used here can, for example, be found in \citet{kneiske:2002a}.
\paragraph{DS spectrum with PHOENIX:}
The atmospheres and spectra of Dark Stars are modeled with the model atmosphere
package PHOENIX, version 16. A basic description of the code can be found in
\citet{hauschildt:1999a} and recent applications to stellar objects can be found in
\citet{short:2009a} and in \citet{fuhrmeister:2010a}. The current
version 16 of the PHOENIX package \citep{hauschildt:2010a} uses the ACES equation
of state (Barman et al., in preparation) to allow for a temperature range from
$100\,$K to above $10^6\,$K and includes a multitude of molecular and dust
species. The models presented here use the 1D mode PHOENIX/1D of PHOENIX with
spherical symmetry. The model atmospheres are computed for
radiative and convective equilibrium with a mixing length to pressure scale
height ratio of 2, with the assumption that no energy is generated
in the atmosphere. The abundance of H was set to 0.92 by number (mass
fraction: 0.75) and that of He was set to 0.08 by number (mass fraction:
0.25) for all models, all other elements (including Li) have an abundance of 
zero in these models. Models with effective temperatures
from $5\,000\,$K to $7\,500\,$K with parameters (gravities, masses) taken from \citet{spolyar:2009a} have been computed.
For all the models a number of different variants by, for
example, varying the line profiles or the equation of state setup to
investigate the effects on the models and the synthetic spectra have been computed. For each
setup discussed below, the individual models are relaxed to equilibrium
before computing high-resolution synthetic spectra. 

\begin{figure*}[tb]
\includegraphics[angle=180,width=\hsize]{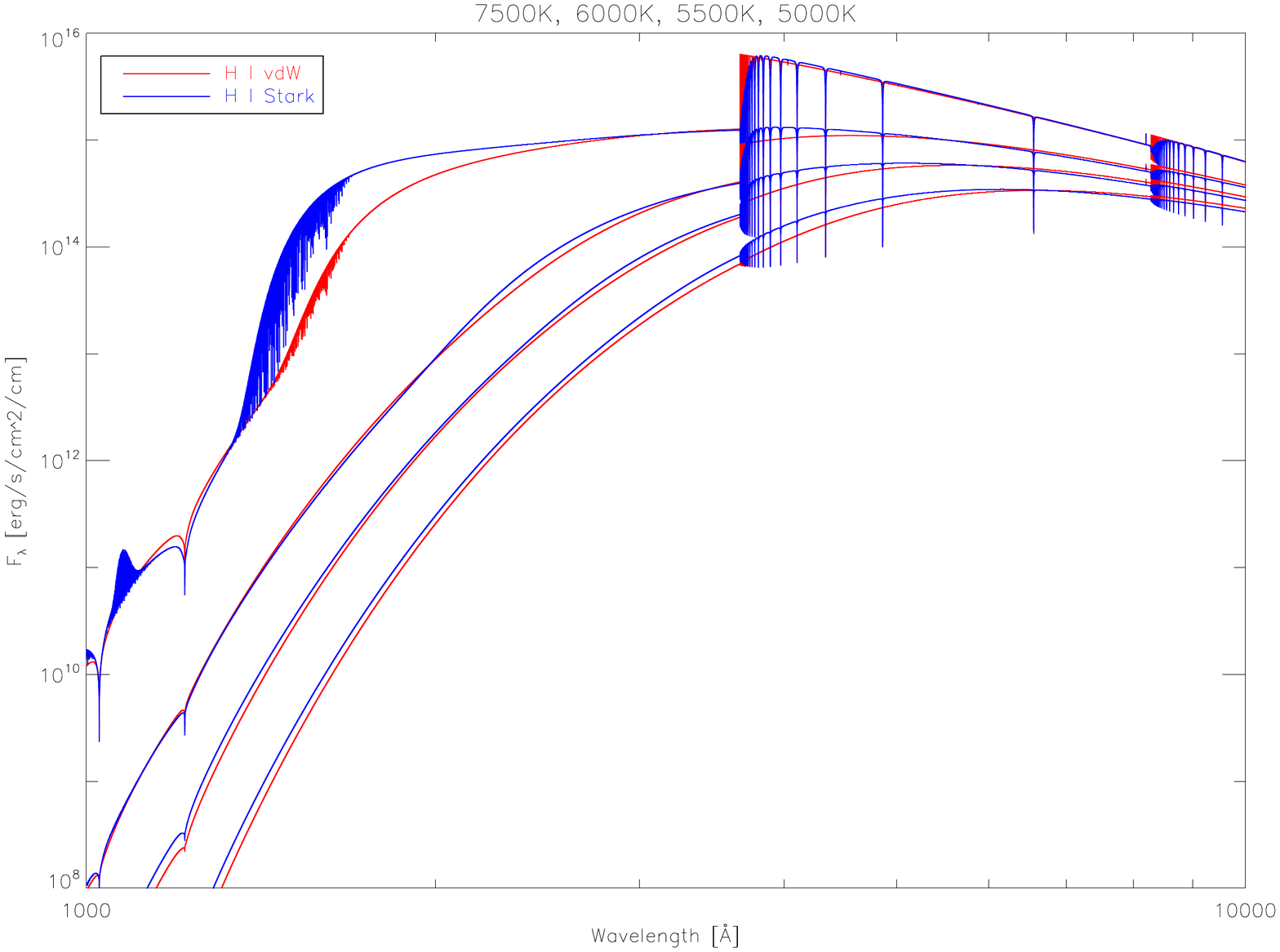}
\caption{Synthetic spectra for DS models derived with the PHOENIX code. The model
parameters are from top to bottom: $\Teff=7\,500\K$, $\logg=1.0$, $M=690\Msun$,
$\Teff = 6\,000\K$, $\logg=0.0$, $M=371\Msun$, $\Teff = 5\,500\K$, $\logg=-0.5$, $M=106\Msun$ and
$\Teff = 5\,000\K$, $\logg=-0.5$, $M=106\Msun$. The gravitational acceleration $g$ is given in CGS units.
      \label{Fig:ProfilesUV}}
\end{figure*}

\begin{figure*}[tb]
\includegraphics[angle=180,width=\hsize]{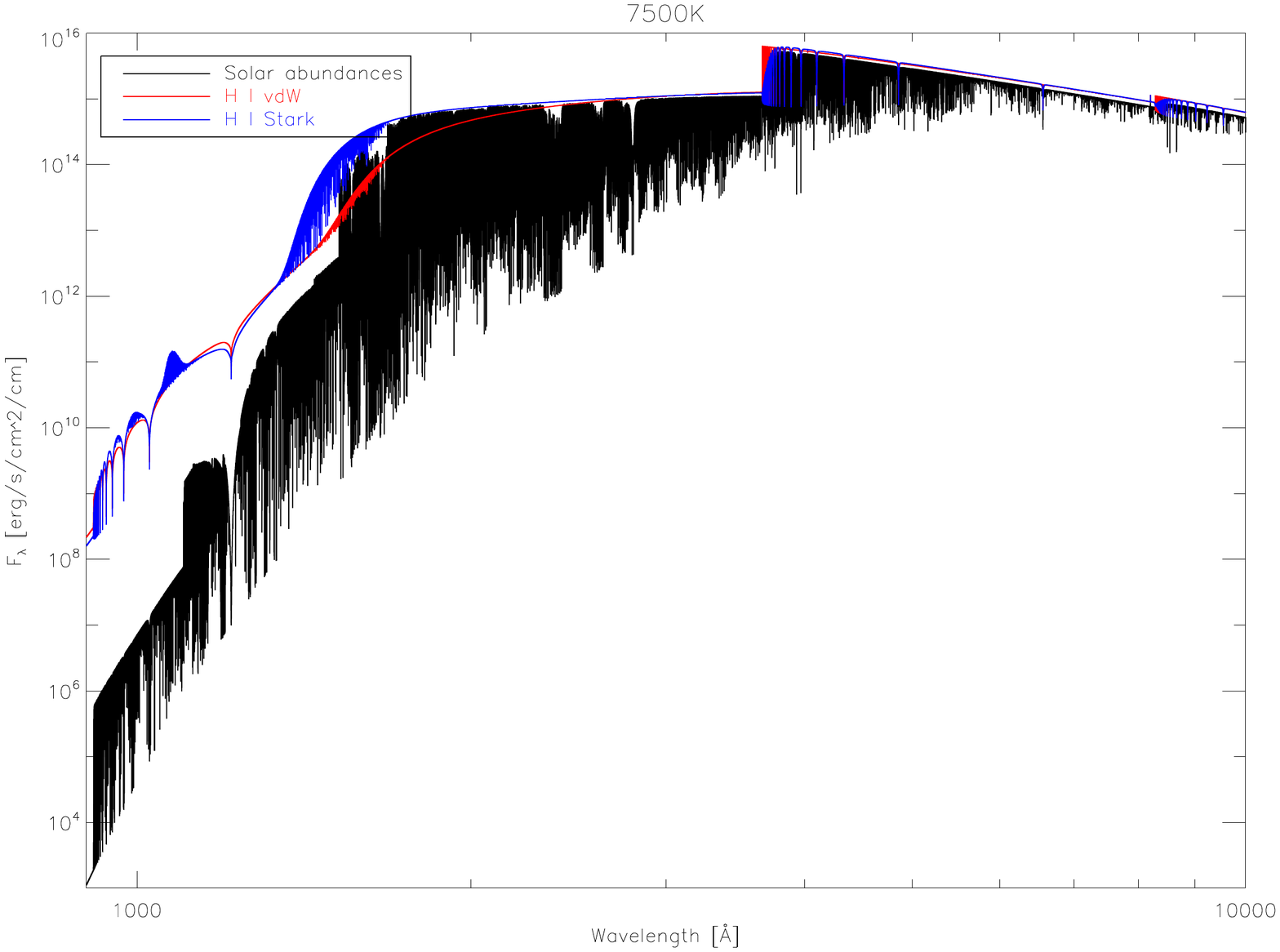}
\caption{Comparison of the Dark Star synthetic spectra for
$\Teff=7\,500\K$, $\logg=1.0$, $M=690\Msun$ with a solar abundances
model with the parameters $\Teff=7\,400\K$, $\logg=1.0$, $M=1\Msun$.
      \label{Fig:CompSolar}}
\end{figure*}

For a given effective temperature, the choice of line profiles produces
the largest variations in the emitted spectra (see Fig.~\ref{Fig:ProfilesUV}).
The two sets of models were computed by (a) using Stark profiles for the 
H lines and (b) using van der Waals (vdW) broadening for the H lines. At 
$\Teff=5\,000\,\K$ the vdW broadening is the dominant line broadening process
whereas at  $\Teff=7\,500\,\K$ Stark broadening is more important, 
despite the absence of free electrons from the light metals, see 
the comparison to a spectrum with solar abundances in
Fig.~\ref{Fig:CompSolar}. The hottest
model in Fig.~\ref{Fig:ProfilesUV} clearly shows t he electronic lines
of H$_2$ in their UV spectra, these are also present in the DS spectra with lower effective
temperatures but are much weaker due to their high excitation energies.
In solar abundance models these are overwhelmed by the metal lines
in the UV and would not be detectable.
For the model parameters considered, the NLTE (Non Local Thermal Equilibrium) effects are small
and appear to be insignificant. At higher 
effective temperatures this will likely be different, however, this parameter
range is not considered here. 

As can be seen from Fig. \ref{Fig:ProfilesUV} no significant hydrogen ionizing radiation is emitted for the stellar temperature 
range considered here.

\paragraph{DS formation rate:}
The formation density of DSs $\dot{n}_{\ast}(z)$ can be linked to the star formation rate $\dot{\rho}_{\ast}(z)$ for the first stars (Pop III).
\citet{spolyar:2008a, spolyar:2009a} propose that every Pop III star that forms inside
the center of an undisturbed dark mater halo 
could establish
a dark star phase 
$\Delta t_{\text{DS}}$. This implies that the mass formation rate of DS can be written as: $\dot{\rho}_{\ast}(z) = \dot{n}_{\ast}(z) \times M_{\text{DS}}$\footnote{This assumption holds true if one specific DS model with mass $M_{\text{DS}}$ is considered.}.
The comoving star formation rate (SFR) of the first stars that form in such environments are calculated by \citet{trenti:2009a} with a variety 
of radiative and chemical (metal enrichment) feedback parameters and for different numbers of stars forming. They obtain values 
for the Pop III mass formation rate of
$10^{-5} - 10^{-3} \unit{M}_{\odot} \unit{year}^{-1} \unit{Mpc}^{-3}$ up to $z=10$ depending on the exact model parameters considered.
Pop III star formation rates in a similar range are also found for lower redshifts \citep{schneider:2006c, tornatore:2007a, maio:2010a}. 
There exist two negative feedback mechanisms for Pop III star formation: radiative feedback from $\unit{H}_2$ ionizing photons and metal enrichment\footnote{For a detailed description of these processes please refer to \citet{trenti:2009a} and references therein.}. 
For DSs, due to their cooler temperatures and longer lifetimes than Pop III stars, both mentioned feedback mechanisms can be suppressed and thus DS formation can, in principle, be enhanced and prolonged.
It is also possible that DS can grow to larger masses than usual first stars ($M_{\text{DS}} > M_{\text{Pop III}}$) which results in a higher SFR $\dot{\rho}_{\ast}(z)$.
In summary it can be concluded that the SFR of DS suffers from large uncertainties. Therefore a reasonable fiducial value is derived from the Pop III SFR ($10^{-5}\unit{M}_{\odot} \unit{year}^{-1} \unit{Mpc}^{-3}$) and a wide range limited by extreme models ($10^{-7} - 10^{-3}\unit{M}_{\odot} \unit{year}^{-1} \unit{Mpc}^{-3}$) is explored.

Hence, as a simplification, a constant star formation rate over a certain redshift period is assumed which can be expressed as a mass formation rate in units of $\unit{M}_{\odot} \unit{year}^{-1} \unit{Mpc}^{-3}$: 
\begin{align}
 \dot\rho_{\ast}(z) = \begin{cases}
                       \unit{SFR}_{\text{Norm}} &\,\text{for }z_{\text{min}} \leq z \leq z_{\text{max}}\\
		       0 &\,\text{else}
                      \end{cases} 
 \mathrm{\label{formula:SFR}}
\end{align}
where $\unit{SFR}_{\text{Norm}}$ is a normalization factor, varied in the above-mentioned range, $z_{\text{min}}$ indicates 
the minimal value of redshift $z$ where Dark Star formation can still occur and $z_{\text{max}}$ denotes the beginning of the
Dark Star formation epoch. The ansatz for the SFR used in this paper is strongly simplified and more elaborate calculations of the SFR are available (see, e.g., \citealt{greif:2006a}, \citealt{sandick:2011a}). \citet{raue:2009a} have investigated the impact of a wide range of SFRs on the resulting EBL density and found the position and overall height of the peak the dominant factor, while the choice of the shape only resulted in a weak change in the resulting peak EBL density (factor $\sim$2). Given the range of the parameter investigated in this work here (e.g., the overall normalization of the SFR ranges over several orders of magnitude), this simplified approach for the SFR is sufficient.

As the duration of the DS-forming period in the universe is directly linked
to the amount of photons that are emitted by Dark Stars the influence of
$z_{\text{min}}$ and $z_{\text{max}}$ on the EBL is also explored. The contribution of $\dot{\rho}_{\ast}(z)$ for large $z$ is suppressed because of the redshift dilution of the photon field which goes as $(1+z)^{-3}$ and so the value of $z_{\text{max}}$ is, in the following, set to 30. For $z_{\text{min}}$ (the end of the Dark Star formation epoch) values 
between 5 and 15 are considered. These values are in good agreement with assumed Pop III formation periods \citep{schneider:2006c, tornatore:2007a, trenti:2009a, trenti:2009b,
maio:2010a}.

\paragraph{DS luminosity:}
Independent of the exact mechanism powering the DM burning, models predict a stable phase which dominates the total radiative output during the DS phase \citep{spolyar:2008a, iocco:2008a}. During this phase the luminosity is nearly constant (see e.g. Figure 2 in \citealt{spolyar:2009a}, Figure 4 in \citealt{iocco:2008a} and Figure 1 in \citealt{yoon:2008a}).
Therefore the following ansatz is adopted for $L_{\nu}$ 
\begin{align}
 L_{\nu}(t(z)-t(z')) = \begin{cases}
                       L_{\nu}^0 &\,\text{for }t(z)-t(z') \leq \Delta t_{\text{DS}} \\
		       0 &\,\text{else}
                      \end{cases} 
\end{align}
with $\Delta t_{\text{DS}}$ being the duration of this stable phase (also referred to as DS lifetime) and $L_{\nu}^0$ being the specific DS luminosity according to its synthetic spectrum (cf. Fig. \ref{Fig:ProfilesUV}). For $\Delta t_{\text{DS}} \ll t(z_{\text{min}}) - t(z_{\text{max}})$ the emissivity calculation reduces to
\begin{equation}
 \varepsilon_{\nu}(z) \approx L^0_{\nu} \, \Delta t_{\text{DS}} \int\limits_z^{z_{\text{max}}}\text{d}z'\,\dot{\rho}_{\ast}(z')
 \text{\label{formula:linemitds}}
\end{equation}
leading to a linear scaling of the resulting EBL with $\Delta t_{\text{DS}}$. 
 The exact length of the DS lifetime is not constrained and depends on various factors, e.g., DM type, DS model, DM halo profile, etc. (for an extensive discussion see \citealt{zackrisson:2010a}). In this work a wide band of possible DS lifetimes is explored, ranging from $10^5$ to $10^9$ years.

The total bolometric luminosity of a single DS is connected to DM particle properties via:
\begin{equation}
 L =\int\text{d}\nu\, L_{\nu}^0 \propto \dfrac{{\langle \sigma v \rangle}_{\text{ann}}}{m_{\chi}}
\end{equation}
\citep{spolyar:2008a}.

Given a certain stellar mass formation rate the DS luminosity produced per mass has to be specified in the model.
The mass range of Pop III stars, and, for this reason, also the mass range of DS is not very well constrained but is expected to be within roughly 10 up to a few hundred solar masses \citep{abel:2002a, schaerer:2002a}. 
The model assumptions for the DM burning and the star formation lead to a wide spread in the DS luminosity per stellar mass.  
This luminosity-to-(stellar)mass ratio (LMR) of published DS models \citep{iocco:2008a, spolyar:2009a, freese:2010a} can be computed and used as input parameters for the EBL calculation presented here 
and range from $\sim 10^2 - 10^5 \unit{L}_{\odot}/\unit{M}_{\odot}$.
 
The influence of DM particle properties on the LMR will be further discussed in Sect. \ref{subsec:DM}.
\section{Results}
\label{sec:results}
\subsection{Constraining the DS parameter space}
\begin{figure}
   \centering
   \includegraphics[width=\hsize]{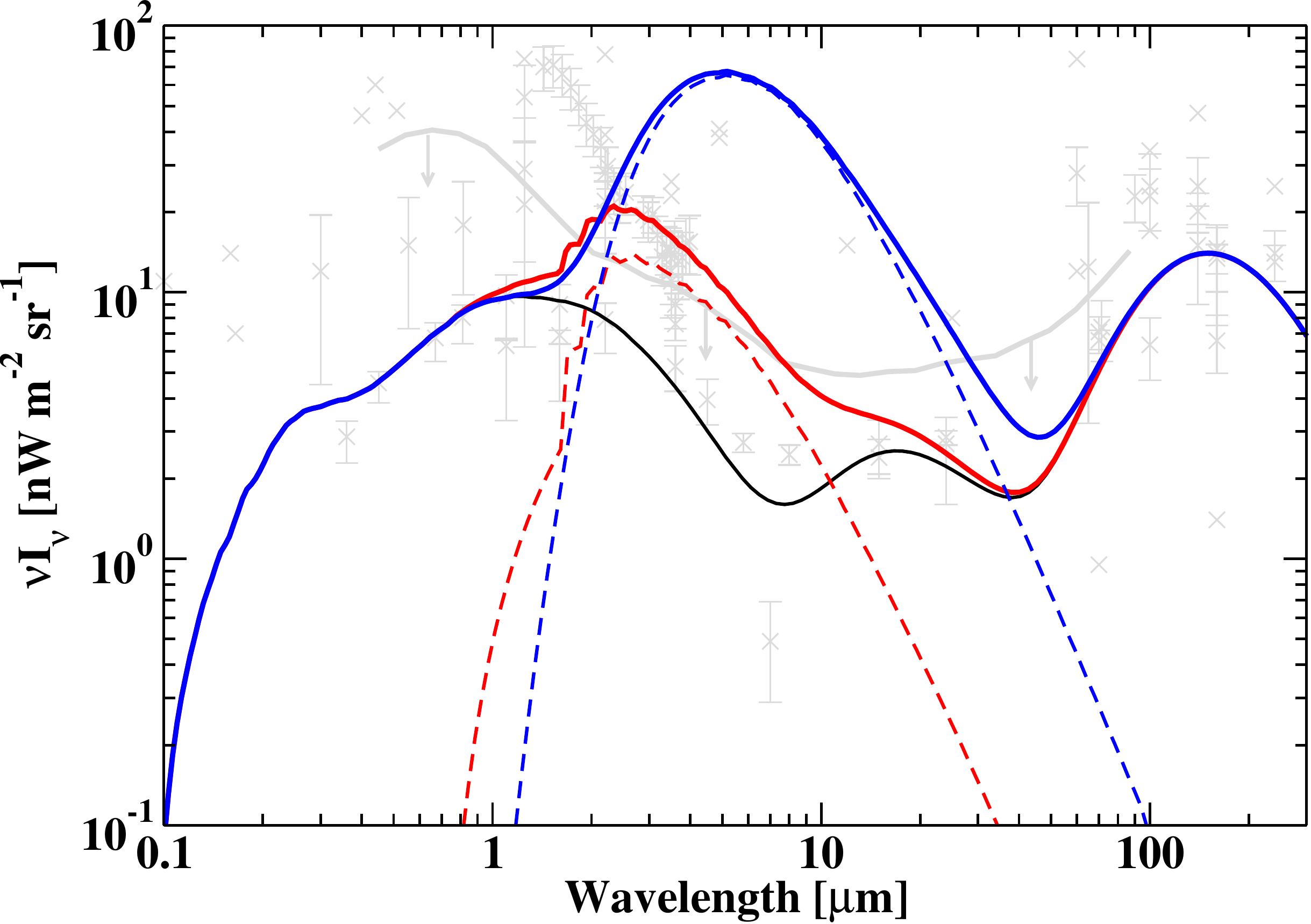}
      \caption{Two different Dark Star parameter sets (red-dashed: $T_{\text{DS}}$ = 7500K, $M = 690$ M$_{\odot}$; blue-dashed: $T_{\text{DS}}$ = 5000 K, $M = 106$ M$_{\odot}$). Both models are calculated with SFR$_{\text{Norm}}$ = 10$^{-3}$, $\Delta t_{\text{DS}}$ = 10$^9$ years, $z_{\text{min}}$ = 5. Grey markers: EBL measurements \& limits adopted from \citet{Mazin:2007a}; grey: upper limits from TeV observations (realistic model) from \citet{Mazin:2007a}. Black: EBL lower limit by \citet{kneiske:2010a}. The total EBL shape in presence of a DS contribution is the sum of the lower limit and the specific DS signature (red and blue lines). 
         \label{FigResults}}
\end{figure}

The EBL contribution of DS is calculated using the methods and parameters discussed in the previous section.
The range of the specific parameter values is shown in table \ref{table:parameters}. A fiducial set of ``intermediate'' parameter values is also displayed which acts as default when a single parameter is varied.
\begin{table}
\begin{center}
\caption{Dark Star parameter range\label{table:parameters}}
\begin{tabular}{ c c c c c c}
\tableline\tableline
 & $\text{L}_{\odot} / \text{M}_{\odot}$ & $\Delta t_{\text{DS}}$ & $z_{\text{min}}$ & $\unit{SFR}_{\text{Norm}}$ \\
\tableline
   min & $10^2$ & $10^5$ & 15 & $10^{-7}$\\      
   fiducial & $10^3$ & $10^7$ & 10 & $10^{-5}$\\
   max & $10^5$ & $10^9$& 5 & $10^{-3}$\\
\tableline
\end{tabular}
\tablecomments{The maximum value of LMR is close to the Eddington limit and therefore acts as a natural upper boundary for the DS parameter space.}
\end{center}
\end{table}
As can be seen from Eqn. \ref{formula:SFR} in combination with Eqn. \ref{formula:EBL} \& Eqn. \ref{formula:emissivity} the resulting EBL density scales linearly with the SFR and the LMR.
In Fig. \ref{FigResults} the EBL contribution for two different DS models is displayed in comparison with a strict lower limit for the guaranteed astrophysical EBL from stars and dust in galaxies \citep{kneiske:2010a}.
Comparing the resulting DS EBL signatures (red dashed and blue dashed curves) with their respective input spectra (see Fig. \ref{Fig:ProfilesUV}) it can be seen that the spectral features are smoothed by the redshift integration.
The peak of the EBL is for both models located at wavelengths $> 2 \mu\text{m}$ which differs from the value ($\lambda \sim 1 \mu\text{m}$) for Pop III stars \citep{santos:2002a} as expected due to lower effective temperatures.
The peak of both EBL signatures clearly reaches into the detectable region of the infrared background. This shows that the EBL offers the potential to constrain DS parameter space.
A DS contribution from these models would result into a total EBL which is the sum of the lower limit and the DS signature (red and blue lines). These EBL shapes are already disfavored by upper limits from TeV observations (grey line).

\begin{figure}
   \centering
   \includegraphics[width=\hsize]{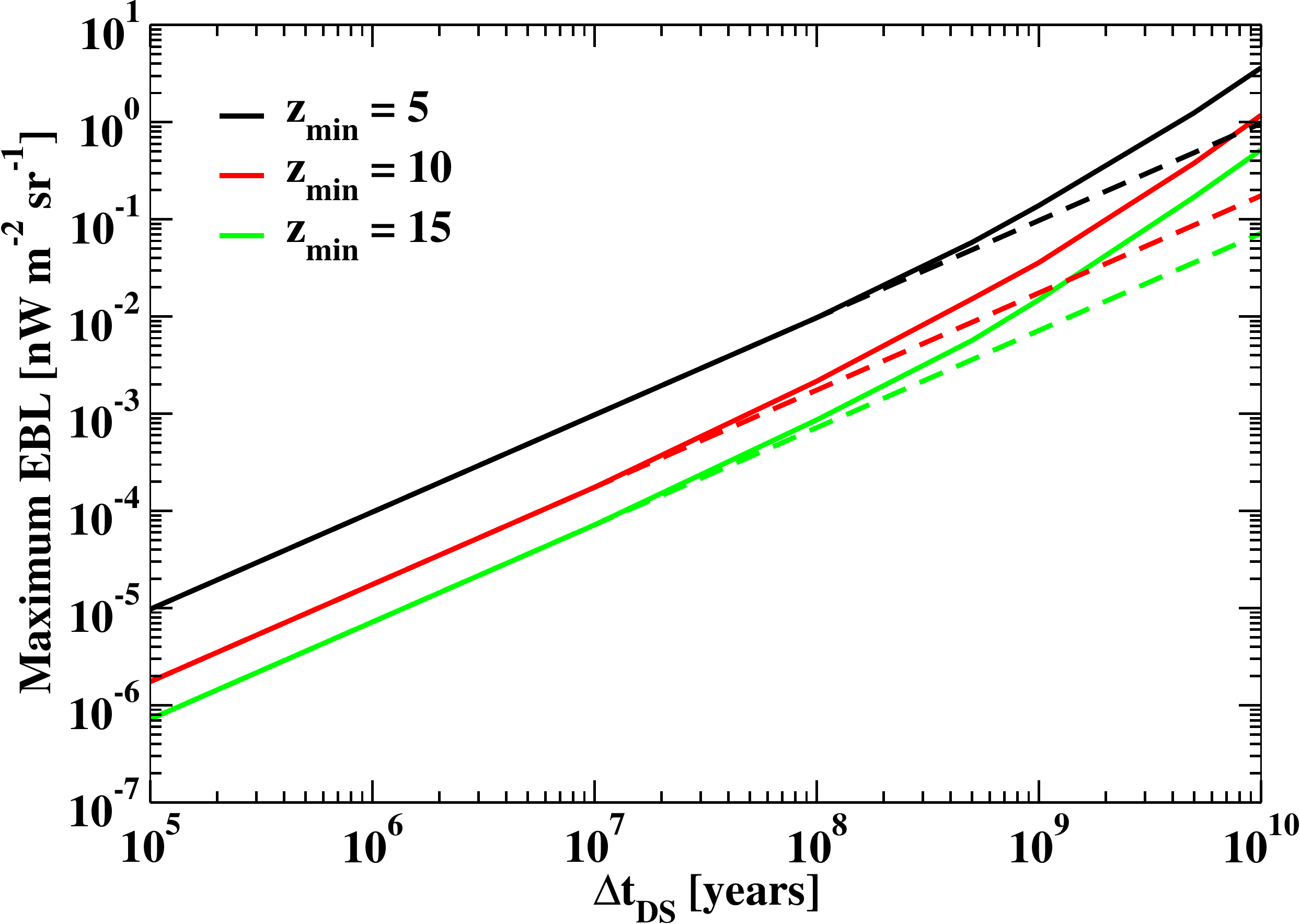}
      \caption{Peak EBL contribution for the $7\,500$ K DS model as a function of DS lifetime, for SFR$_{\text{Norm}}$ the fiducial value of $10^{-5}$ is used. 
         \label{FigPeakEbl}}
   \end{figure}

\begin{figure}[tb]
   \centering
   \includegraphics[width=\hsize]{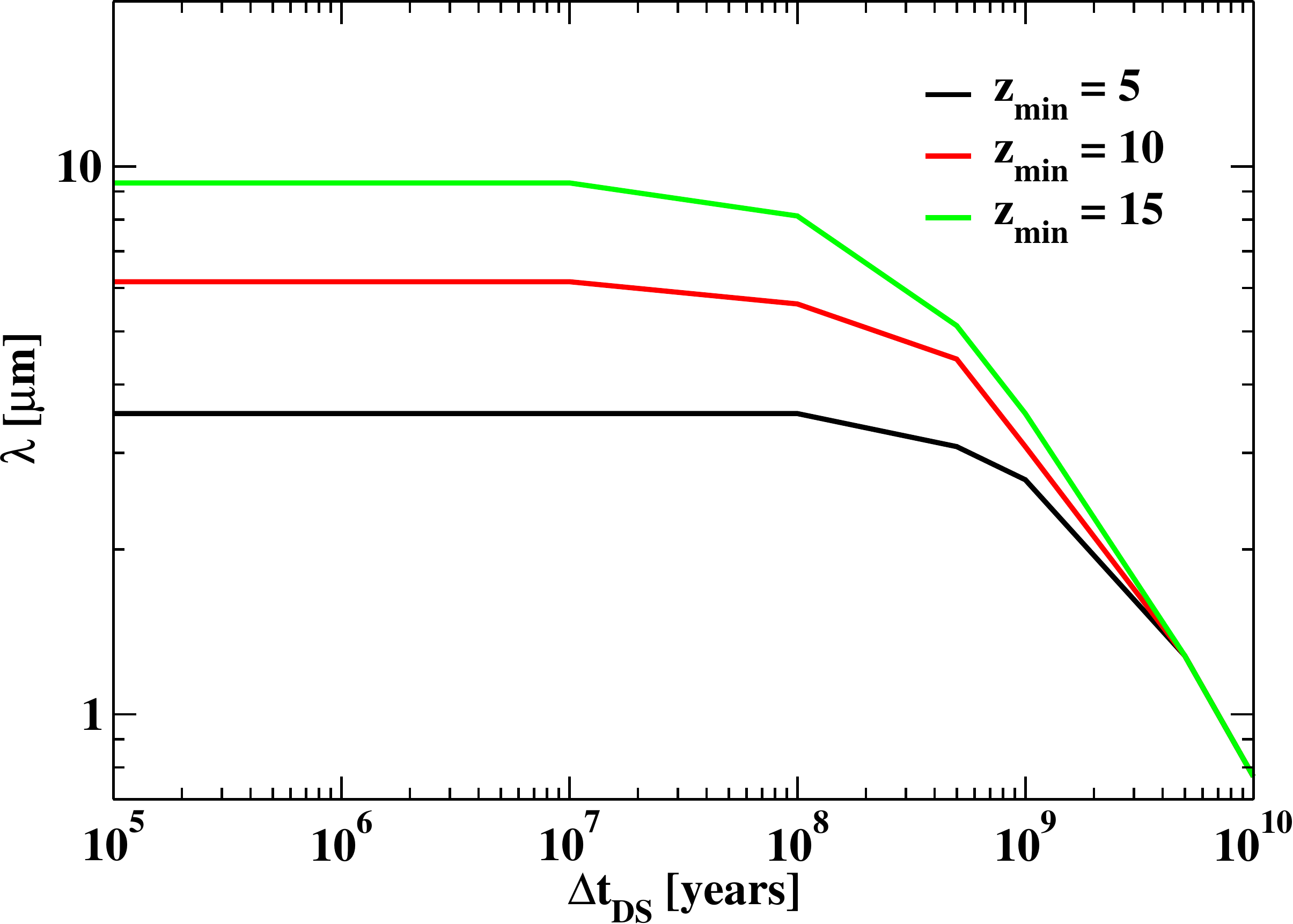}
      \caption{Location wavelength of the peak in the EBL SED as a function of DS lifetime for the $7\,500 \text{ K}$ DS model. For $\Delta t_{\text{DS}} > 10^9 \text{ years}$ the values converge toward the intrinsic emission maximum indicating that DSs still emit light at $z = 0$. 
         \label{FigPeakWl}}
   \end{figure}

In Fig. \ref{FigPeakEbl} the peak EBL contribution for three values of $z_{\text{min}}$ with respect to varying DS lifetimes is displayed. 
As a consequence of Eqn. \ref{formula:linemitds}, for DS lifetimes smaller than the formation period $t(z_{\text{min}}) - t(z_{\text{max}})$ the resulting EBL scales linearly with increasing $\Delta t_{\text{DS}}$. At higher lifetimes than $\sim 10^8 \unit{years}$ the intensity of the EBL is increased to a greater amount as well as the peak value of the DS signature is shifted toward lower wavelengths (Fig. \ref{FigPeakWl}). This is caused by a residual emissivity at lower redshifts $z < z_{\text{min}}$ as the end of DS formation is not the end of DS photon emission. Due to the strong dilution of the photon number density with redshift $(1+z)^{-3}$ the most recent emission dominates the EBL contribution. If $\Delta t_{\text{DS}}$ is short enough the end of DS formation is roughly equal to the end of DS emitting photons as the amount of DS drops of almost instantly. The dashed lines display a linear relationship between $\Delta t_{\text{DS}}$ and the maximum EBL flux. Please note that DS with lifetimes $\Delta t_{\text{DS}} \simeq 10^{10}$ years, as displayed in Fig. \ref{FigPeakEbl} and Fig. \ref{FigPeakWl}, would be still present in today's universe and therefore most likely to be detected.

The model results can be summarized in a formula giving the peak EBL contribution at $z=0$ from a DS population normalized to the fiducial DS parameters.
\begin{align}
  (\nu I_{\nu})_{\max} = & 2\times10^{-5} \text{ nW m}^{-2} \text{ sr}^{-1} \times\left(\frac{\Delta t_{\text{DS}}}{10^7 \unit{years}}\right) \nonumber \\ & \times\left(\frac{\text{SFR}_{\text{Norm}}}{10^{-5}}\right) 
 \times\left(\frac{\text{LMR}}{10^3 \unit{L}_{\odot}/\unit{M}_{\odot}}\right) \times\left(\frac{z_{\text{min}}}{10}\right)^{-2.5}
 \text{\label{formula:EBLformula}}
\end{align}
The calculation of the resulting EBL contribution via this formula offers a conservative estimate as the DS lifetime only enters linearly which is true for $\Delta t_{\text{DS}}$ up to values as large as $\sim 10^8$ years. The possible enhancement of the EBL contribution due to longer DS lifetimes is not taken into account here, but one can estimate it from Fig. \ref{FigPeakEbl} or it has to be calculated as described in this work.
Comparing lower with upper limits on the EBL one finds an allowed EBL contribution from DS in the range of $5 - 25 \text{ nW m}^{-2} \text{ sr}^{-1}$ for wavelength between $2$ and $10 \mu \text{m}$ (see e.g. \citealt{Mazin:2007a} Fig. 15). Adopting this range limits on DS parameters can be derived. For example: A DS with $M_{\text{DS}} = 106 \text{ M}_{\odot}$, $L_{\text{DS}} = 9 \times 10^6 \text{ L}_{\odot}$, a DS lifetime $\Delta t_{\text{DS}} = 10^8 \text{ years}$ and minimum formation redshift $z_{\text{min}} = 5$ results in a constraint on the DS formation rate between $5 \times 10^{-4}$ and $3 \times 10^{-3} \unit{M}_{\odot} \unit{year}^{-1} \unit{Mpc}^{-3}$. In this way several DS parameter sets can be used to constrain specific DS scenarios.

\subsection{Implications for dark matter properties}
\label{subsec:DM}
The dominant factor for the DS luminosity due to DM burning is the square of the DM density distribution $\rho_{\chi}$ which is determined by the original DM density profile inside primordial virialized halos, the change of the profile under the influence of contracting baryonic material (adiabatic contraction), and repopulation of the inner cusp of the DM halo due to scattering processes \citep{freese:2008c, freese:2009b}. 
The characteristics (temperature, radius, luminosity) of a DS rely on the efficiency of the DM luminosity compared to other energy generation mechanism like nuclear fusion. In general the more DM contributes to the total stellar luminosity of the star, the cooler and bigger it grows regardless of the mechanism that produces the high DM density inside the star.

As explained above the DS luminosity scales with $\rho^2_{\chi}$. The replenishment of the DM density inside a DS is possible either by the gravitational pull of the collapsing gas onto the DS or by multiple elastic scattering processes between DM and the stellar baryons or a combination of both. The LMR of the specific dark matter powered stars presented in \citet{spolyar:2009a} tend to be slightly dependent on ${\langle \sigma v \rangle}_{\text{ann}} / m_{\chi}$; higher values result in higher DS masses but at the same time also to higher DS luminosities. For the early stages of DS formation the LMR is nearly proportional to $\sqrt{{\langle \sigma v \rangle}_{\text{ann}} / m_{\chi}}$. In the case of DM capture, as investigated by \citet{iocco:2008a}, the DS luminosity scales linearly with the elastic scattering cross section $\sigma_0$. It is also shown that DM burning due to WIMP capture is more efficient in low-mass stars, but under the assumption that the product $\sigma_0 \rho_{\chi}$ is of the order $10^{-26} \text{ GeV} \text{ cm}^{-1}$ \citet{yoon:2008a} have shown that a great mass range of stars can undergo a stable DM burning phase.

It is further worthwhile to note that, while conservative models of the Dark Star contribution to the EBL do not strongly constrain WIMP properties such as mass or annihilation cross section, this picture can change when our prior assumptions are relaxed: As outlined in \citet{spolyar:2009a}, the Dark Star luminosity-to-mass ratio - and therefore the peak EBL contribution (see Eqn. \ref{formula:EBLformula}) - will roughly scale with the square root of the annihilation cross section and WIMP mass as $\text{LMR} \propto \sqrt{\langle \sigma v \rangle}_{\text{ann} / m_{\chi}}$ where $m_{\chi}$ is the mass of the WIMP. Assuming cross-sections of the order $10^{-23} \text{cm}^{3} \text{s}^{-1}$ for TeV scale WIMPs, as have recently been invoked to explain PAMELA measurements of the cosmic-ray positron fraction \citep{cirelli:2008a}, at least the "max" model presented in Table 1 would already be strongly constrained by existing EBL data. In comparison to other indirect dark matter detection channels, the Dark Star induced EBL component is unique in that it is not sensitive to the exact branching ratios of the annihilation yields into photons or charged particles and their resulting spectra, as all annihilation products (with the exception of neutrinos, of course) are trapped and thermalized within the Dark Star. In this regard, the calculations presented here can also serve as an independent and complementary template for indirect dark matter searches in other wave bands.

Implications also exist regarding the elastic scattering cross section of the Dark Matter particles. As capture via elastic scattering is a dominant channel for replenishing the fuel of the Dark Star (see e.g. \citealt{iocco:2008a}), once equilibrium has been reached the luminosity of such a star scales linearly with the WIMP - proton scattering cross-section\footnote{In the case of extended adiabatic contraction being the main mechanism of DM replenishment the DS luminosity scales only with the stellar mass \citep{freese:2010a}.} (hydrogen of course being the dominant target for an assumed primordial mixture). There is considerable discussion ongoing with regard to experimental searches for elastic WIMP - nucleon scattering in underground experiments.

With respect to Dark Stars, it is interesting that, again assuming the basic setup from \citet{iocco:2008a}, a spin-dependent WIMP-proton cross section of the order 1pb for a WIMP with a rest mass of tens of GeV would already result in Dark Stars with LMRs very close to the ``max'' value of table \ref{table:parameters}. While backing up the constraints down to the multiple-femtobarn level from the non-detection of annihilation neutrinos from the Sun \citep{hooper:2008a}, the as of yet definitive missing detection of a Dark Star contribution to the EBL therefore reinforces the conclusion that spin-dependent scattering can reconcile the DAMA and CoGENT measurements of an annual modulation of event rates \citep{bernabei:2010a, aalseth:2010a} with null-results from other experiments only for cases in which the WIMP-neutron cross-section is several orders of magnitude greater than the WIMP-proton cross section, as e.g. discussed in \citet{ullio:2001a}. It should however be noted that for the case of light Dark Matter particles, there exist windows where a surprisingly high spin-independent scattering cross section of the order $10^{-40} \unit{cm}^2$ is compatible with the DAMA/LIBRA and CoGENT measurements \citep{hooper:2011a}. This may again be a case where Dark Matter annihilation might significantly effect the life cycle of the first stars, and should be a focus of future work.
\section{Discussion \& Conclusion}
\label{sec:discussion}
WIMP dark matter can have an impact on the evolution of the first stars. In this work, the detectability of DS generated signatures in the EBL is investigated. This approach opens a new window to search for DM effects: the near-infrared (NIR). The EBL contributions from different DS parameter sets have been calculated and, for certain sets of parameters, the resulting EBL flux can reach into the detectable range of the infrared background. A parameterization is presented to calculate the peak EBL contribution for a variety of model input values. The resulting EBL can be used to constrain certain DS scenarios by comparing the calculated peak EBL contribution of DS with existing upper limits in the EBL density. 
The results of this work can be seen as complementary approach to put constraints on DS formation scenarios as investigated by \citet{schleicher:2009a} where the influence of DS to reionization, the $\gamma$-ray and neutrino background has been investigated.

\citet{freese:2010a} recently proposed a scenario in which Dark Stars can reach
enormously high masses ($10^7 \text{ M}_{\odot}$) due to very efficient and long-lasting accretion processes of 
baryonic matter onto the Dark Star, so-called super massive Dark Stars (SMDS). Under these conditions very long lived and luminous SMDS can be produced. As suggested by the authors nearly all baryonic matter inside the DM halo the SMDS forms in is accreted onto the star. This fact translates in a very high SFR which can be used as input for Eqn. \ref{formula:EBLformula} to put also constraints on the SMDS scenarios.

As a final note, it should be stressed that the ability to constrain DS scenarios via EBL depends crucially on the improvement of EBL limits and measurements. The upcoming James Webb Space Telescope will provide new data on deep field galaxy counts, as well as refined direct measurements of the infrared background. The next generation of Imaging Air Cherenkov Telescopes, like the Cherenkov Telescope Array \citep{cta:2010a}, will radically improve upon existing TeV measurements, hopefully providing for a direct measurement of the EBL - induced cutoff in AGN spectra (see e.g. \citealt{raue:2010a}). Fully self-consistent models of Dark Star formation, internal structure and evolution are also demanding to finally tackle the question whether WIMP annihilation played a role in the first luminous objects in the universe.

\acknowledgements
This work was made possible with the support of the Cluster of Excellence: ``Connecting Particles with the Cosmos'', part of the Landesexzellenzinitiative Hamburg, and the
collaborative research center (SFB) 676 ``Particle, Strings and the early
Universe'' at the University of Hamburg. The authors like to thank Alessandro Mirizzi and G\"otz Heinzelmann for reading of the manuscript and helpful comments. The authors also like to thank the anonymous referee for useful comments.


\begin{thebibliography}{52}
\expandafter\ifx\csname natexlab\endcsname\relax\def\natexlab#1{#1}\fi

\bibitem[{{Aalseth} {et~al.}(2011){Aalseth}, {Barbeau}, {Bowden},
  {Cabrera-Palmer}, {Colaresi}, {Collar}, {Dazeley}, {de Lurgio}, {Fast},
  {Fields}, {Greenberg}, {Hossbach}, {Keillor}, {Kephart}, {Marino}, {Miley},
  {Miller}, {Orrell}, {Radford}, {Reyna}, {Tench}, {van Wechel}, {Wilkerson},
  \& {Yocum}}]{aalseth:2010a}
{Aalseth}, C.~E., {et~al.} 2011, Physical Review Letters, 106, 131301

\bibitem[{{Abazajian} {et~al.}(2009){Abazajian}, {Adelman-McCarthy},
  {Ag{\"u}eros}, {Allam}, {Allende Prieto}, {An}, {Anderson}, {Anderson},
  {Annis}, {Bahcall}, \& et~al.}]{abazajian:2009a}
{Abazajian}, K.~N., {et~al.} 2009, \apjs, 182, 543

\bibitem[{{Abel} {et~al.}(2002){Abel}, {Bryan}, \& {Norman}}]{abel:2002a}
{Abel}, T., {Bryan}, G.~L., \& {Norman}, M.~L. 2002, Science, 295, 93

\bibitem[{{Barman} {et~al.}(2011){Barman}, {Macintosh}, {Konopacky}, \&
  {Marois}}]{barman:2011a}
{Barman}, T.~S., {Macintosh}, B., {Konopacky}, Q.~M., \& {Marois}, C. 2011,
  \apj, 733, 65

\bibitem[{{Bernabei} {et~al.}(2010){Bernabei}, {Belli}, {Cappella}, {Cerulli},
  {Dai}, {D'Angelo}, {He}, {Incicchitti}, {Kuang}, {Ma}, {Montecchia},
  {Nozzoli}, {Prosperi}, {Sheng}, {Wang}, \& {Ye}}]{bernabei:2010a}
{Bernabei}, R., {et~al.} 2010, European Physical Journal C, 67, 39

\bibitem[{{Bertone} {et~al.}(2005){Bertone}, {Hooper}, \&
  {Silk}}]{bertone:2005a}
{Bertone}, G., {Hooper}, D., \& {Silk}, J. 2005, \physrep, 405, 279

\bibitem[{{Bromm} \& {Larson}(2004)}]{bromm:2004a}
{Bromm}, V., \& {Larson}, R.~B. 2004, \araa, 42, 79

\bibitem[{{Cirelli} \& {Strumia}(2008)}]{cirelli:2008a}
{Cirelli}, M., \& {Strumia}, A. 2008, ArXiv e-prints: 0808.3867

\bibitem[{{Colless} {et~al.}(2003){Colless}, {Peterson}, {Jackson}, {Peacock},
  {Cole}, {Norberg}, {Baldry}, {Baugh}, {Bland-Hawthorn}, {Bridges}, {Cannon},
  {Collins}, {Couch}, {Cross}, {Dalton}, {De Propris}, {Driver}, {Efstathiou},
  {Ellis}, {Frenk}, {Glazebrook}, {Lahav}, {Lewis}, {Lumsden}, {Maddox},
  {Madgwick}, {Sutherland}, \& {Taylor}}]{colless:2003a}
{Colless}, M., {et~al.} 2003, ArXiv Astrophysics e-prints: astro-ph/0306581

\bibitem[{{Dwek} {et~al.}(1998){Dwek}, {Arendt}, {Hauser}, {Fixsen}, {Kelsall},
  {Leisawitz}, {Pei}, {Wright}, {Mather}, {Moseley}, {Odegard}, {Shafer},
  {Silverberg}, \& {Weiland}}]{dwek:1998a}
{Dwek}, E., {et~al.} 1998, \apj, 508, 106

\bibitem[{{Fazio} {et~al.}(2004){Fazio}, {Ashby}, {Barmby}, {Hora}, {Huang},
  {Pahre}, {Wang}, {Willner}, {Arendt}, {Moseley}, {Brodwin}, {Eisenhardt},
  {Stern}, {Tollestrup}, \& {Wright}}]{fazio:2004a}
{Fazio}, G.~G., {et~al.} 2004, \apjs, 154, 39

\bibitem[{{Finke} {et~al.}(2010){Finke}, {Razzaque}, \& {Dermer}}]{finke:2010a}
{Finke}, J.~D., {Razzaque}, S., \& {Dermer}, C.~D. 2010, \apj, 712, 238

\bibitem[{{Franceschini} {et~al.}(2008){Franceschini}, {Rodighiero}, \&
  {Vaccari}}]{franceschini:2008a}
{Franceschini}, A., {Rodighiero}, G., \& {Vaccari}, M. 2008, \aap, 487, 837

\bibitem[{{Freese} {et~al.}(2009){Freese}, {Gondolo}, {Sellwood}, \&
  {Spolyar}}]{freese:2009b}
{Freese}, K., {Gondolo}, P., {Sellwood}, J.~A., \& {Spolyar}, D. 2009, \apj,
  693, 1563

\bibitem[{{Freese} {et~al.}(2010){Freese}, {Ilie}, {Spolyar}, {Valluri}, \&
  {Bodenheimer}}]{freese:2010a}
{Freese}, K., {Ilie}, C., {Spolyar}, D., {Valluri}, M., \& {Bodenheimer}, P.
  2010, \apj, 716, 1397

\bibitem[{{Freese} {et~al.}(2008){Freese}, {Spolyar}, \&
  {Aguirre}}]{freese:2008c}
{Freese}, K., {Spolyar}, D., \& {Aguirre}, A. 2008, Journal of Cosmology and
  Astro-Particle Physics, 11, 14

\bibitem[{{Freese} {et~al.}(2008{\natexlab{a}}){Freese}, {Bodenheimer},
  {Spolyar}, \& {Gondolo}}]{freese:2008b}
{Freese}, K., {Bodenheimer}, P., {Spolyar}, D., \& {Gondolo}, P.
  2008{\natexlab{a}}, \apjl, 685, L101

\bibitem[{{Fuhrmeister} {et~al.}(2010){Fuhrmeister}, {Schmitt}, \&
  {Hauschildt}}]{fuhrmeister:2010a}
{Fuhrmeister}, B., {Schmitt}, J.~H.~M.~M., \& {Hauschildt}, P.~H. 2010, \aap,
  511, A83+

\bibitem[{{Gardner} {et~al.}(2006){Gardner}, {Mather}, {Clampin}, {Doyon},
  {Greenhouse}, {Hammel}, {Hutchings}, {Jakobsen}, {Lilly}, {Long}, {Lunine},
  {McCaughrean}, {Mountain}, {Nella}, {Rieke}, {Rieke}, {Rix}, {Smith},
  {Sonneborn}, {Stiavelli}, {Stockman}, {Windhorst}, \&
  {Wright}}]{gardner:2006a}
{Gardner}, J.~P., {et~al.} 2006, Space Science Reviews, 123, 485

\bibitem[{{Gilmore} {et~al.}(2009){Gilmore}, {Madau}, {Primack}, {Somerville},
  \& {Haardt}}]{gilmore:2009a}
{Gilmore}, R.~C., {Madau}, P., {Primack}, J.~R., {Somerville}, R.~S., \&
  {Haardt}, F. 2009, \mnras, 399, 1694

\bibitem[{{Greif} \& {Bromm}(2006)}]{greif:2006a}
{Greif}, T.~H., \& {Bromm}, V. 2006, \mnras, 373, 128

\bibitem[{Hauschildt \& Baron(1999)}]{hauschildt:1999a}
Hauschildt, P.~H., \& Baron, E. 1999, Journal of Computational and Applied
  Mathematics, 109, 41

\bibitem[{{Hauschildt} \& {Baron}(2010)}]{hauschildt:2010a}
{Hauschildt}, P.~H., \& {Baron}, E. 2010, \aap, 509, A36+

\bibitem[{{Hauser} {et~al.}(1998){Hauser}, {Arendt}, {Kelsall}, {Dwek},
  {Odegard}, {et~al.}}]{hauser:1998a}
{Hauser}, M.~G., {Arendt}, R.~G., {Kelsall}, T., {Dwek}, E., {Odegard}, N.,
  {et~al.} 1998, The Astrophysical Journal, 508, 25

\bibitem[{{Hauser} \& {Dwek}(2001)}]{hauser:2001a}
{Hauser}, M.~G., \& {Dwek}, E. 2001, \araa, 39, 249

\bibitem[{{Hooper} \& {Kelso}(2011)}]{hooper:2011a}
{Hooper}, D., \& {Kelso}, C. 2011, \prd, 84, 083001

\bibitem[{{Hooper} {et~al.}(2009){Hooper}, {Petriello}, {Zurek}, \&
  {Kamionkowski}}]{hooper:2008a}
{Hooper}, D., {Petriello}, F., {Zurek}, K.~M., \& {Kamionkowski}, M. 2009,
  \prd, 79, 015010

\bibitem[{{Iocco} {et~al.}(2008){Iocco}, {Bressan}, {Ripamonti}, {Schneider},
  {Ferrara}, \& {Marigo}}]{iocco:2008a}
{Iocco}, F., {Bressan}, A., {Ripamonti}, E., {Schneider}, R., {Ferrara}, A., \&
  {Marigo}, P. 2008, \mnras, 390, 1655

\bibitem[{{Iocco}(2008)}]{iocco:2008b}
{Iocco}, F. 2008, \apjl, 677, L1

\bibitem[{Jungman {et~al.}(1996)Jungman, Kamionkowski, \&
  Griest}]{jungman:1996a}
Jungman, G., Kamionkowski, M., \& Griest, K. 1996, Physics Reports, 267, 195

\bibitem[{{Kneiske} {et~al.}(2004){Kneiske}, {Bretz}, {Mannheim}, \&
  {Hartmann}}]{kneiske:2004a}
{Kneiske}, T.~M., {Bretz}, T., {Mannheim}, K., \& {Hartmann}, D.~H. 2004, \aap,
  413, 807

\bibitem[{{Kneiske} \& {Dole}(2010)}]{kneiske:2010a}
{Kneiske}, T.~M., \& {Dole}, H. 2010, \aap, 515, A19+

\bibitem[{{Kneiske} {et~al.}(2002){Kneiske}, {Mannheim}, \&
  {Hartmann}}]{kneiske:2002a}
{Kneiske}, T.~M., {Mannheim}, K., \& {Hartmann}, D.~H. 2002, \aap, 386, 1

\bibitem[{{Komatsu} {et~al.}(2011){Komatsu}, {Smith}, {Dunkley}, {Bennett},
  {Gold}, {Hinshaw}, {Jarosik}, {Larson}, {Nolta}, {Page}, {Spergel},
  {Halpern}, {Hill}, {Kogut}, {Limon}, {Meyer}, {Odegard}, {Tucker}, {Weiland},
  {Wollack}, \& {Wright}}]{komatsu:2011a}
{Komatsu}, E., {et~al.} 2011, \apjs, 192, 18

\bibitem[{{Madau} \& {Pozzetti}(2000)}]{madau:2000a}
{Madau}, P., \& {Pozzetti}, L. 2000, \mnras, 312, L9

\bibitem[{{Maio} {et~al.}(2010){Maio}, {Ciardi}, {Dolag}, {Tornatore}, \&
  {Khochfar}}]{maio:2010a}
{Maio}, U., {Ciardi}, B., {Dolag}, K., {Tornatore}, L., \& {Khochfar}, S. 2010,
  \mnras, 905

\bibitem[{{Mazin} \& {Raue}(2007)}]{Mazin:2007a}
{Mazin}, D., \& {Raue}, M. 2007, \aap, 471, 439

\bibitem[{{Peebles}(1993)}]{peebles:1993a}
{Peebles}, P.~J.~E. 1993, {Principles of physical cosmology} (Princeton Series
  in Physics, Princeton, NJ: Princeton University Press, |c1993)

\bibitem[{{Primack} {et~al.}(2008){Primack}, {Gilmore}, \&
  {Somerville}}]{primack:2008a}
{Primack}, J.~R., {Gilmore}, R.~C., \& {Somerville}, R.~S. 2008, in American
  Institute of Physics Conference Series, Vol. 1085, American Institute of
  Physics Conference Series, ed. F.~A. {Aharonian}, W.~{Hofmann}, \&
  F.~{Rieger}, 71--82

\bibitem[{{Raue} {et~al.}(2009){Raue}, {Kneiske}, \& {Mazin}}]{raue:2009a}
{Raue}, M., {Kneiske}, T., \& {Mazin}, D. 2009, \aap, 498, 25

\bibitem[{{Raue} \& {Mazin}(2010)}]{raue:2010a}
{Raue}, M., \& {Mazin}, D. 2010, Astroparticle Physics, 34, 245

\bibitem[{{Salamon} \& {Stecker}(1998)}]{salamon:1998a}
{Salamon}, M.~H., \& {Stecker}, F.~W. 1998, \apj, 493, 547

\bibitem[{{Sandick} {et~al.}(2011){Sandick}, {Diemand}, {Freese}, \&
  {Spolyar}}]{sandick:2011a}
{Sandick}, P., {Diemand}, J., {Freese}, K., \& {Spolyar}, D. 2011, \jcap, 1, 18

\bibitem[{{Santos} {et~al.}(2002){Santos}, {Bromm}, \&
  {Kamionkowski}}]{santos:2002a}
{Santos}, M.~R., {Bromm}, V., \& {Kamionkowski}, M. 2002, \mnras, 336, 1082

\bibitem[{{Schaerer}(2002)}]{schaerer:2002a}
{Schaerer}, D. 2002, \aap, 382, 28

\bibitem[{{Schleicher} {et~al.}(2009){Schleicher}, {Banerjee}, \&
  {Klessen}}]{schleicher:2009a}
{Schleicher}, D.~R.~G., {Banerjee}, R., \& {Klessen}, R.~S. 2009, \prd, 79,
  043510

\bibitem[{{Schneider} {et~al.}(2006){Schneider}, {Salvaterra}, {Ferrara}, \&
  {Ciardi}}]{schneider:2006c}
{Schneider}, R., {Salvaterra}, R., {Ferrara}, A., \& {Ciardi}, B. 2006, \mnras,
  369, 825

\bibitem[{{Short} \& {Hauschildt}(2009)}]{short:2009a}
{Short}, C.~I., \& {Hauschildt}, P.~H. 2009, \apj, 691, 1634

\bibitem[{{Spolyar} {et~al.}(2009){Spolyar}, {Bodenheimer}, {Freese}, \&
  {Gondolo}}]{spolyar:2009a}
{Spolyar}, D., {Bodenheimer}, P., {Freese}, K., \& {Gondolo}, P. 2009, \apj,
  705, 1031

\bibitem[{{Spolyar} {et~al.}(2008){Spolyar}, {Freese}, \&
  {Gondolo}}]{spolyar:2008a}
{Spolyar}, D., {Freese}, K., \& {Gondolo}, P. 2008, Physical Review Letters,
  100, 051101

\bibitem[{{Springel} {et~al.}(2005){Springel}, {White}, {Jenkins}, {Frenk},
  {Yoshida}, {Gao}, {Navarro}, {Thacker}, {Croton}, {Helly}, {Peacock}, {Cole},
  {Thomas}, {Couchman}, {Evrard}, {Colberg}, \& {Pearce}}]{springel:2005a}
{Springel}, V., {et~al.} 2005, \nat, 435, 629

\bibitem[{{Stecker} {et~al.}(2006){Stecker}, {Malkan}, \&
  {Scully}}]{stecker:2006a}
{Stecker}, F.~W., {Malkan}, M.~A., \& {Scully}, S.~T. 2006, \apj, 648, 774

\bibitem[{{The CTA Consortium}(2010)}]{cta:2010a}
{The CTA Consortium}. 2010, ArXiv e-prints: 1008.3703

\bibitem[{{Tornatore} {et~al.}(2007){Tornatore}, {Ferrara}, \&
  {Schneider}}]{tornatore:2007a}
{Tornatore}, L., {Ferrara}, A., \& {Schneider}, R. 2007, \mnras, 382, 945

\bibitem[{{Trenti} \& {Stiavelli}(2009)}]{trenti:2009a}
{Trenti}, M., \& {Stiavelli}, M. 2009, \apj, 694, 879

\bibitem[{{Trenti} {et~al.}(2009){Trenti}, {Stiavelli}, \& {Michael
  Shull}}]{trenti:2009b}
{Trenti}, M., {Stiavelli}, M., \& {Michael Shull}, J. 2009, \apj, 700, 1672

\bibitem[{{Ullio} {et~al.}(2001){Ullio}, {Kamionkowski}, \&
  {Vogel}}]{ullio:2001a}
{Ullio}, P., {Kamionkowski}, M., \& {Vogel}, P. 2001, Journal of High Energy
  Physics, 7, 44

\bibitem[{{Yoon} {et~al.}(2008){Yoon}, {Iocco}, \& {Akiyama}}]{yoon:2008a}
{Yoon}, S., {Iocco}, F., \& {Akiyama}, S. 2008, \apjl, 688, L1

\bibitem[{{Zackrisson} {et~al.}(2010){Zackrisson}, {Scott}, {Rydberg}, {Iocco},
  {Edvardsson}, {{\"O}stlin}, {Sivertsson}, {Zitrin}, {Broadhurst}, \&
  {Gondolo}}]{zackrisson:2010a}
{Zackrisson}, E., {et~al.} 2010, \apj, 717, 257

\end{thebibliography}
\end{document}